# Self-calibrating d-scan: measuring ultrashort laser pulses on-target using an arbitrary pulse compressor

Benjamín Alonso,[1,2,*] Íñigo J. Sola,[1] and Helder Crespo[2,1]

[1]Grupo de Investigación en Aplicaciones del Láser y Fotónica, Departamento de Física Aplicada, University of Salamanca, Salamanca, E-37008, Spain
[2]IFIMUP-IN and Departamento de Física e Astronomia, Universidade do Porto, Rua do Campo Alegre 687, 4169-007, Porto, Portugal
* Corresponding author: b.alonso@usal.es

## ABSTRACT

In most applications of ultrashort pulse lasers, temporal compressors are used to achieve a desired pulse duration in a target or sample, and precise temporal characterization is important. The dispersion-scan (d-scan) pulse characterization technique usually involves using glass wedges to impart variable, well-defined amounts of dispersion to the pulses, while measuring the spectrum of a nonlinear signal produced by those pulses. This works very well for broadband few-cycle pulses, but longer, narrower bandwidth pulses are much more difficult to measure this way. Here we demonstrate the concept of *self-calibrating d-scan*, which extends the applicability of the d-scan technique to pulses of arbitrary duration, enabling their complete measurement without prior knowledge of the introduced dispersion. In particular, we show that the pulse compressors already employed in chirped pulse amplification (CPA) systems can be used to simultaneously compress and measure the temporal profile of the output pulses on-target in a simple way, without the need of additional diagnostics or calibrations, while at the same time calibrating the often-unknown differential dispersion of the compressor itself. We demonstrate the technique through simulations and experiments under known conditions. Finally, we apply it to the measurement and compression of 27.5 fs pulses from a CPA laser.

## Introduction

Ultrashort laser pulses are a key tool for many applications in science and technology, from materials processing to time-resolved spectroscopy and attosecond science, with wavelength ranges spanning from the THz to the extreme ultraviolet and beyond. Commonly used sources include mode-locked laser oscillators, laser amplifiers, optical parametric oscillators/amplifiers and pulse post-compressors. These sources are usually coupled to a temporal pulse compressor, either to obtain a compressed pulse or other pulse duration (or chirp) on target. For example, achieving maximum compression (and hence maximum signal) at the focus of a nonlinear microscope requires pre-compensating the dispersion introduced by all the optics within the system (e.g. filters, polarizers, and objectives) between the laser source and the sample plane, which invariably requires an adequate pulse compressor. Temporal pulse compressors are also an integral part of a very important class of ultrafast sources, namely chirped-pulse amplification (CPA) laser systems and optical parametric chirped pulse amplification (OPCPA) systems.

Knowing the temporal profile of the pulses on target is very important for optimizing the compression and for establishing the pulse quality and duration, which, in turn, determines the attainable peak power and temporal resolution in time-resolved experiments. Since the advent of ultrashort laser pulses, the community of users has been working in the development of methods to measure such short events, most of them based on nonlinear effects involving the pulse itself or a comparable ultrashort pulse[1]. Some of these techniques are well-established and present many variants.

Recently, the dispersion-scan (d-scan) technique was introduced[2,3], which originally consisted in the measurement of the second-harmonic generation (SHG) and/or sum-frequency generation (SFG) signal produced by a pulse in a nonlinear medium for different and *well-known* values of dispersion introduced by a chirped mirror and glass wedge compressor around the maximum compression point. The resulting measurement, known as the d-scan trace, encodes information about the spectral phase of the pulses. By coupling this measurement with a suitable mathematical model and optimization algorithm, one is able to fully retrieve the spectral phase of the pulses from the d-scan trace without the need of approximations. The d-scan technique has several advantages, including a robust inline experimental setup devoid of temporal delay lines, beam splitting or beam recombination, as well as relaxed SHG conversion bandwidth requirements compared to other techniques, as demonstrated in Ref. 3. Furthermore, d-scan is highly robust to noise and can correctly retrieve the spectral phase even when there is no SHG signal for a given frequency, as long as there is SFG signal resulting from the mixing of that frequency with other parts of the pulse spectrum. Additionally, a d-scan setup also doubles as a pulse compressor, which together with its measurement capability allows one to simultaneously measure the pulse and to optimize its compression.

Until now, the variable dispersion has mostly been introduced with glass wedges of known material and angle. By controlling the steps of the wedge insertion, the amount of glass crossed by the pulse is easily determined and its dispersion can be obtained from the corresponding refractive index calculated, e.g., from Sellmeier equations. This information is used within the optimization algorithm that calculates the spectral phase of the pulse, which is expressed as a multi-parameter unknown variable[2]. This implementation and approach has enabled, e.g., the generation and measurement of sub-two-cycle[4] and even near-single-cycle[5-8] amplified laser pulses post-compressed in a hollow-core fibre (HCF). On the other hand, there is also strong interest in measuring longer, multi-cycle pulses, as directly produced by Ti:Sapphire CPA laser systems and by OPCPAs, since this is determinant for many applications involving those pulses, such as materials processing, ultrafast spectroscopy, pulse post-compression, high-harmonic generation and laser particle acceleration experiments, among others. However, the relatively narrower bandwidth of these pulses, compared to e.g., post-compressed pulses, makes it very difficult in practice to introduce enough dispersion with standard glass wedges to obtain an adequate d-scan trace.

On the other hand, the variable pulse compressors used in ultrafast CPA systems, which usually employ diffraction gratings, prisms, grisms or combinations of some of these elements, are capable of introducing much larger dispersion than glass wedges and are designed to compress the output pulses of the specific laser system they are part of, providing optimized dispersion and minimum pulse duration for a specific position of the compressor. However, the compressors in CPA laser systems are usually not calibrated, in the sense that neither their exact dispersion at a given position nor their (differential) dispersion change per step are known to arbitrary order, since this is not a necessary requirement for their operation.

In this work, we present a self-calibrating dispersion-scan technique capable of measuring and compressing ultrashort laser pulses over a broad range of pulse parameters, as produced by a broad range of ultrafast sources, and where prior knowledge of the amount of dispersion introduced for each position or step of the compressor is not required. The reconstruction algorithm is capable of retrieving not only the ultrashort pulse, but also of obtaining the differential dispersion introduced during the scan (hence the term *self-calibrating dispersion-scan*), which is an important property with great practical advantages with respect to previous works[9-11]. As a consequence, d-scan measurements can be performed with virtually any dispersion scanning system. For instance, the non-calibrated internal pulse compressors in CPA laser systems can be used to directly measure their output pulses, which effectively enables extending the d-scan technique to longer, narrower bandwidth laser pulses, with durations from tens of femtoseconds up to several picoseconds and more, while maintaining the distinctive advantages of the technique.

## Results and Discussion

### Theoretical demonstration of self-calibrating d-scan with ~30-fs pulses.

The self-calibrating d-scan operation (detailed in the Methods Section) is based on the numerical retrieval of the spectral phase of the pulses using a nonlinear optimization algorithm, where the spectral phase is treated as a multi-parameter unknown variable, and where the unknown dispersion of the dispersion scanning system or compressor is described by a theoretical model of its functional dependence on the compressor position or step. The fundamental spectrum of the pulse can be measured directly or retrieved from the measured d-scan trace[2,7]. During numerical optimization, the d-scan trace is initially simulated for a random guess phase and then this phase is iteratively modified until the simulated d-scan trace converges to the experimental one using a merit function based on the comparison between both traces. If the intensity calibrated fundamental spectrum is independently measured, there is no need to calibrate the intensity of the nonlinear signal (e.g., SHG) used for the d-scan trace, since we can use the dispersion marginal of the trace or a generalized wavelength-dependent error function to obtain this calibration[2,6]. In the present work, our optimization algorithm retrieves not only the spectral phase of the pulse but also the unknown phase introduced by the compressor during the scan. For the pulse, each point of its frequency-dependent phase is treated as an independent variable, which enables measuring an arbitrary spectral phase. For the compressor dispersion, the phase is described by a Taylor series truncated to a given arbitrary order. The pulse and its d-scan trace are not affected by constant and linear spectral phase terms, so these two terms are not considered. Actually, for many relevant situations, we can model the dispersion introduced per step as the combination of only two unknown parameters: the group delay dispersion (GDD) and the 3rd-order dispersion (TOD). For example, the prism compressor used in the final part of this work will be modeled with GDD and TOD, in agreement with previous works[12,13]. Please note that these approximations refer to the variable part of the compressor dispersion, which tends to be well-behaved (or even monotonic, as in the case of a chirped mirror and wedge compressor), and not to the spectral phase of the pulse itself, which can be arbitrarily complex. If necessary, more terms can be added (e.g., 4th-order dispersion). The compressor dispersion can be also described by a complex function of frequency, but this is usually not necessary in most practical real-world situations, as demonstrated further below.

We start by presenting the self-calibrating d-scan method through numerical simulations assuming 28 fs pulses, from which we retrieve both the compressor dispersion (Fig. 1) and the pulse measurement (Fig. 2). Such pulse durations are usually beyond the range of a chirped mirror and wedge compressor, so the scan must rely on another type of compressor, such as a grating or a prism compressor. Then, we demonstrate the technique through experiments with broadband pulses using glass wedges of known dispersion and ~7 fs pulses and compare it to the standard d-scan technique. Finally, we apply it to the measurement of pulses with approximately 28 fs from a CPA laser system.

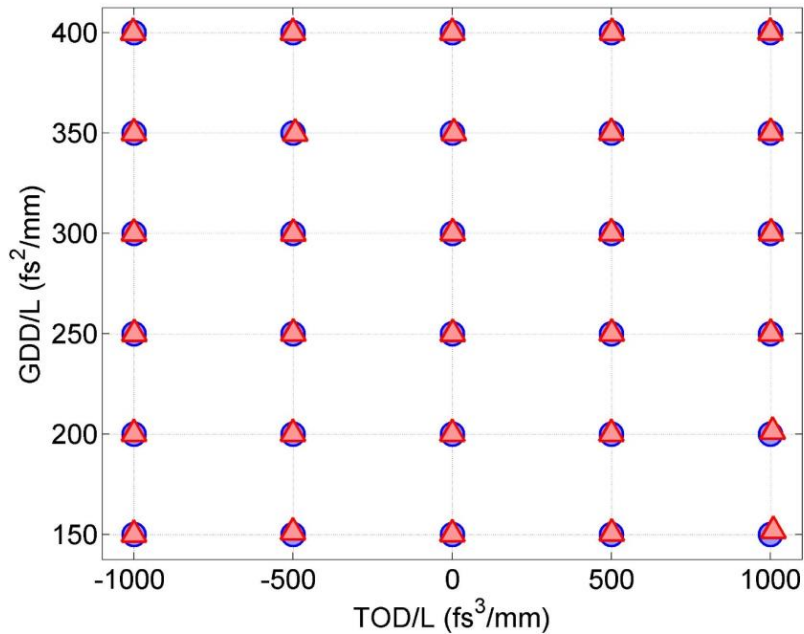

**Figure 1**. **Retrieval of the dispersion introduced in the simulations**. Comparison between the GDD/L and TOD/L values in the simulated and retrieved d-scans. Blue circles: simulated values; red triangles: retrieved values.

As a first test set (30 simulations) we calculated d-scan traces for simulated ultrashort pulses using the same spectral resolution for the SHG signal as in the CPA laser pulse measurements presented further below, and used a measured fundamental spectrum from the same system. Also, we used 51 values of prism insertion per scan, as in those experiments. The spectral phase imparted on the simulated initial pulse was a combination of GDD = 200 fs$^2$, TOD = +5000 fs$^3$, plus an oscillatory term given by $0.25 \cos[100(\omega - \omega_0) + \pi/10]$, where $\omega$ is the frequency and $\omega_0$ is the central frequency of the laser pulse. These parameters result in a theoretical trace [Fig. 2(a)] that is very representative of actual systems.

In the simulations, for the compressor dispersion we used the following values of GDD/L and TOD/L (referring to the dispersion of the grating or prism compressor per unit insertion length): GDD/L = 150, 200, 250, 300, 350, and 400 fs$^2$/mm; TOD/L = -1000, -500, 0, 500, and 1000 fs$^3$/mm. Using always the same simulated pulse described before, and for each combination of GDD/L and TOD/L, we calculated the corresponding d-scan trace, which produced the set of 30 simulations. We then used the self-calibrating retrieval, which gave us the spectral phase as well as the GDD/L and TOD/L for each case. In all the retrievals, the initial guess pulse phase was flat (Fourier-transform limited pulse) and the guess GDD/L and TOD/L were 250 fs$^2$/mm and 0 fs$^3$/mm, respectively. In Fig. 1, we show the comparison between the values used in the simulation and the values obtained with the self-calibrating d-scan algorithm. The achieved agreement demonstrates that it is possible to apply this scheme to compressors with unknown dispersion at least up to the 3$^{rd}$ order. As an example, we show in Fig. 2(a) the theoretical d-scan trace created to test the numerical retrieval for simulation number 15 (GDD/L = 250 fs$^2$/mm and TOD/L = 1000 fs$^3$/mm). The retrieved self-calibrating d-scan trace is plotted in Fig. 2(b) to illustrate the achieved convergence and agreement.

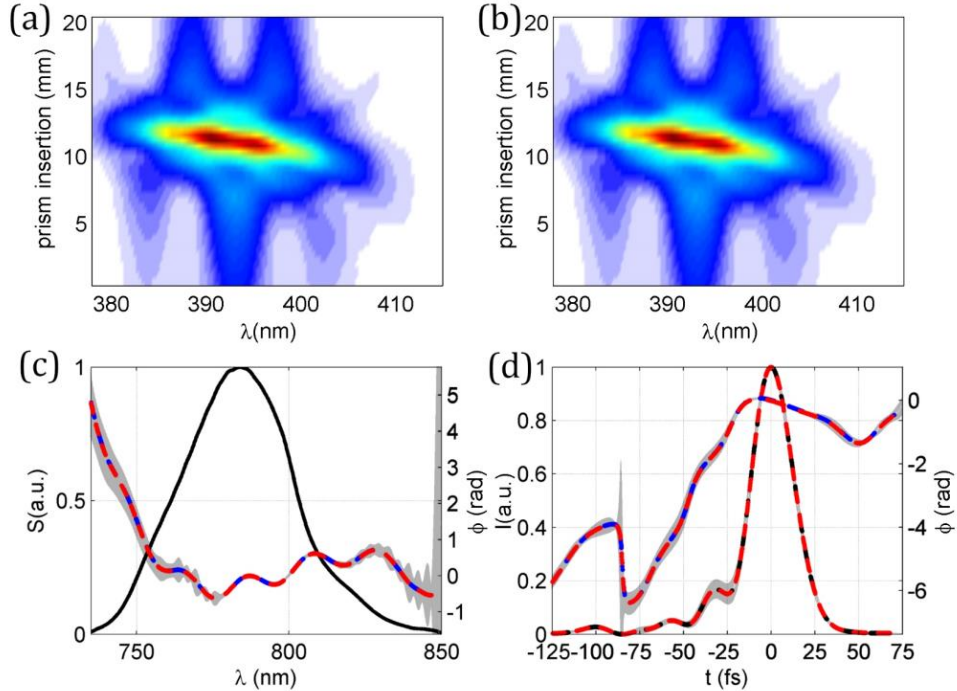

**Figure 2**. **Example and statistics of the simulated self-calibrating d-scan retrievals**. (**a**) Theoretical and (**b**) retrieved self-calibrating d-scan traces for GDD/L = 250 fs$^2$/mm and TOD/L = 1000 fs$^3$/mm (case 15 in Fig. 1). Statistics of the results of the 30 simulations with different GDD and TOD given in Fig. 1: (**c**) pulse spectrum (solid black curve) and retrieved phase (dashed blue curve: theoretical; dashed red curve: mean), (**d**) temporal intensity (dashed black curve: theoretical; dashed red curve: mean), and retrieved phase (dashed blue curve: theoretical; dashed red curve: mean). Grey shaded areas denote the error (magnified 10 times) in the retrievals, calculated as the standard deviation over the 30 retrievals.

We then analyzed the capability of our proposed scheme to retrieve the unknown pulse. The calculated spectral phase matches the theoretical one, as shown in Fig. 2(c). The same stands for the comparison between the theoretical and the retrieved intensity and phase in the temporal domain [Fig. 2(d)], which indicates that the pulse is accurately retrieved. Also, the obtained pulse duration of 28.0±0.1 fs (FWHM) is in excellent agreement with the theoretical pulse duration of 28.0 fs.

The statistical analysis was done over the 30 retrievals obtained for the different differential dispersions shown in Fig. 1, with the error calculated as the standard deviations of the retrieved functions (spectral phase and temporal intensity and phase) for each value of the independent variable (wavelength and time, respectively). In the plots, this error is magnified 10 times in order to be distinguishable. The low values of these errors are indicative of the precision of the retrieval in these simulations.

## Experimental demonstration and comparison with standard d-scan: retrieval of the pulse and of the compressor dispersion.

In a first experimental test, we used a Ti:Sapphire few-cycle ultrafast laser oscillator with a central wavelength of around 800 nm and a Fourier limit of approximately 7 fs, including chirped mirrors and glass wedges to compress the pulse and perform the d-scan, as done previously[2,14]. The d-scan trace was measured by recording the SHG spectra generated by the output beam focused on a BBO crystal (10 µm thick, cut for type I SHG) using a fibre-coupled spectrometer (HR4000, Ocean Optics Inc.) as a function of the dispersion introduced by fixed insertion steps of BK7 glass wedges (8° angle), which was calculated from Sellmeier equations. This allows comparing the results from the standard d-scan algorithm[2] with the present self-calibrating d-scan, where we have modeled the dispersion of BK7 by fitting to orders 2-3 (GDD and TOD) and orders 2-4 (up to 4th-order dispersion) of the corresponding Taylor expansion.

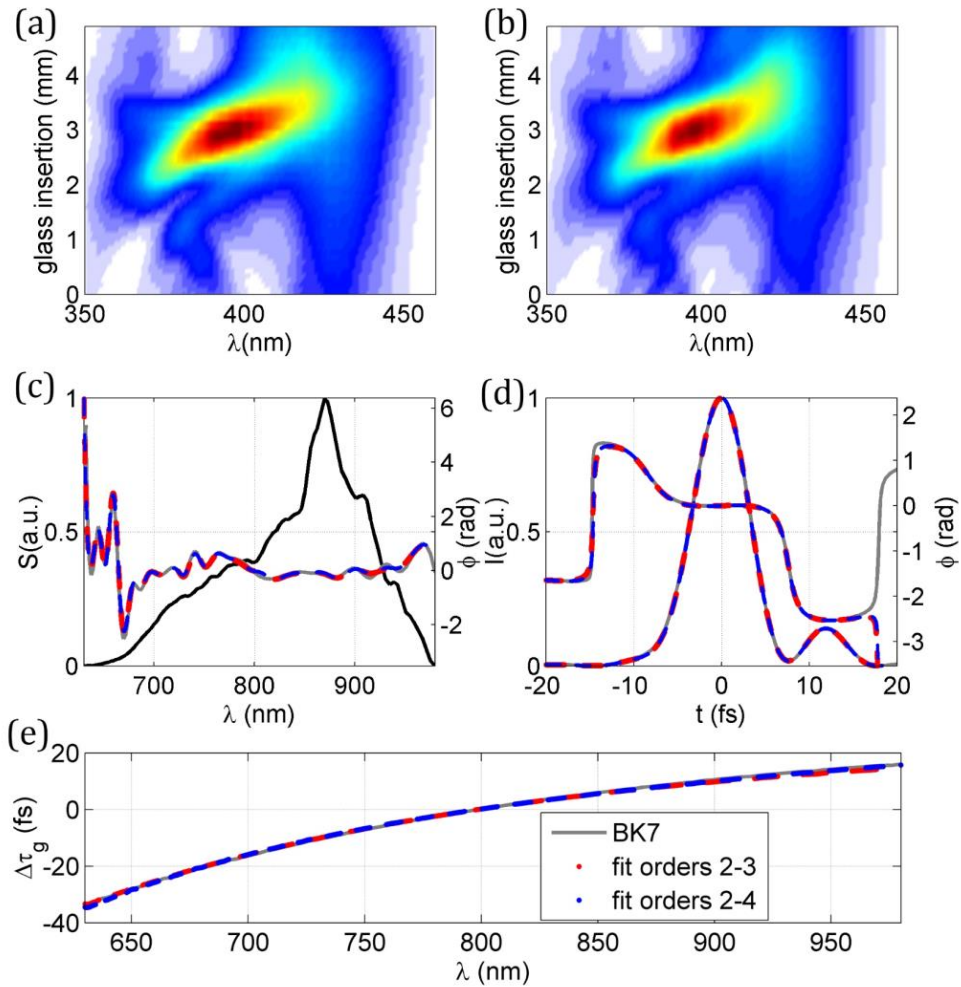

**Figure 3**. **Results of the self-calibrating d-scan of the few-cycle laser oscillator**. (**a**) Experimental and (**b**) retrieved d-scan trace using the self-calibrating d-scan algorithm with glass dispersion fitted up to 3$^{rd}$ order. (**c**) Spectral intensity and retrieved phase, (**d**) retrieved temporal intensity and phase. In (**c**) and (**d**): using the known

BK7 dispersion and the standard d-scan algorithm (solid grey curve); using self-calibrating d-scan with dispersion fit to orders 2-3 (dashed red curve) and orders 2-4 (dashed blue curve). (**e**) Comparison of the variation of the group delay per mm calculated for BK7 (solid grey) and obtained from self-calibrating d-scan using dispersion fit to orders 2-3 (dashed red) and orders 2-4 (dashed blue).

The results for the pulse retrieval are shown in Fig. 3, where the experimental trace in Fig. 3(a) is compared with the self-calibrating trace in Fig. 3(b). The pulse is retrieved for the material insertion corresponding to the optimum compression (the shortest output pulse). For the three cases shown (known dispersion; fit to orders 2-3; fit to orders 2-4), the calculated spectral phases show a very good agreement [Fig. 3(c)], as do the pulse temporal intensity and phase [Fig. 3(d)]. The retrieved pulse durations are, respectively, 7.4, 7.3 and 7.3 fs (FWHM). To complete this first experimental demonstration of the validity of the self-calibrating method, we compared the calibration of the dispersion obtained by the algorithm with the values for BK7 calculated from Sellmeier equations. Since the linear phase term is irrelevant here, we calculated the variation of the group delay, $\tau_g$, per mm, referred to the central wavelength of 800 nm. These results are given in Fig. 3(e), where a very good agreement is observed.

## Application of self-calibrating d-scan to a real CPA laser system.

As stated previously, the self-calibrating d-scan technique can be applied to any CPA or OPCPA laser system, by using its internal (and in most cases uncalibrated) compressor as the dispersion scanning module. Therefore, the laser system can be used to perform a self-diagnostic, where both the output pulses and the compressor calibration can be simultaneously obtained. To demonstrate this, we applied the technique to a CPA Ti:Sapphire laser system, comprising a double-pass compressor with two Brewster-angled prism pairs, delivering pulses with a Fourier-transform limit of 22 fs (FWHM) centered at 785 nm (Femtolasers Femtopower Compact Pro CEP).

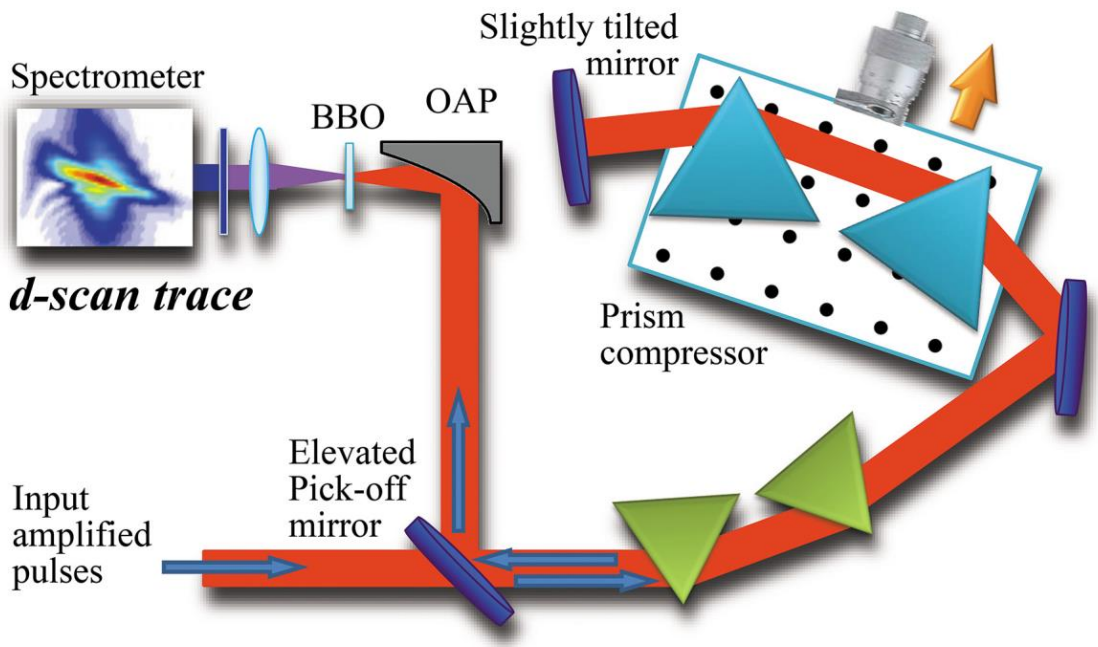

**Figure 4**. **Experimental setup for the prism-compressor self-calibrating d-scan**. The dispersion is scanned from negative to positive by controlling the prism insertion with a translation stage in the second prism pair of the double-pass compressor. The output pulse is focused with an off-axis parabola (OAP) onto a BBO crystal, after which it is collimated with a fused silica lens and the fundamental beam is suppressed with a blue filter. The SHG signal is measured with a spectrometer for 51 consecutive prism insertions, hence producing the d-scan trace.

The dispersion is scanned by varying the prism insertion in the beam path (Fig. 4). The Brewster-angled prisms are made of LaKN16 glass, have an apex angle of approximately 60°, are separated by approximately 2 m and are aligned for minimum deviation. After the prism compressor, the laser output was focused by a gold-coated off-axis parabola (OAP) onto a nonlinear crystal (BBO, 10 µm thick, cut for type I SHG). The SHG signal was filtered with high-pass optical filters (Schott BG filters) and acquired with a fibre-coupled spectrometer (S2000, Ocean Optics Inc.). Additionally, we recorded the spectrum of the fundamental pulse with a second spectrometer (PC2000, Ocean Optics Inc.). For the d-scan measurement, we varied the prism insertion along a range of 20 mm with a linear stage in steps of 0.4 mm; less (more) prism insertion means more (less) negative chirp. This way, we produced a complete d-scan trace from 51 values of prism insertion, in which we tracked the SHG signal from negatively to positively chirped pulses, going through optimum compression. We also ensured there were no nonlinear effects inside the prism compressor (see the Methods Section). Under these conditions, we took five experimental d-scan traces for which the spectral phase, GDD/L and TOD/L were all retrieved in the manner explained previously. The experimental and retrieved d-scan traces are shown in Figs. 5(a) and 5(b), respectively, for one of the experimental measurements.

The statistics of the compressor dispersion calculated from the optimization algorithm are GDD/L = 277.5 ± 2.5 $fs^2$/mm and TOD/L = 175 ± 15 $fs^3$/mm at 785 nm, so both parameters are obtained with reasonable precision. The absolute errors are calculated from the standard deviation of the 5 independent measurements and retrievals, which is the same data set used to calculate the errors in the pulse retrieval in the spectral and temporal domains (the standard deviation is calculated for the 5 retrieved values of the corresponding functions at each wavelength and time, respectively).

In order to ascertain the accuracy of the retrieved differential GDD and TOD of our compressor, we have estimated these parameters using analytical expressions for the phase introduced by a four-prism compressor as a function of the inter-prism distances and insertion[15], also taking into account the propagation in the prisms. For our prism material (LaK16 glass) and geometry (design wavelength of 800 nm; distance between prism pairs of 205 cm) we obtained values of GDD/L = 273 $fs^2$/mm and TOD/L = 187 $fs^3$/mm at 785 nm, in very good agreement with the dispersion retrieved by the self-calibrating d-scan algorithm.

In Fig. 5(c), we show the measured spectrum and the retrieved spectral phase of the pulse, including its standard deviation. Like in the simulations presented before, the deviation is small and only becomes significant at the spectral edges, as expected, so it does not affect the temporal pulse intensity. The spectral phase of the pulse is given for the prism insertion that corresponds to the shortest pulse duration and presents a remaining negative cubic phase term, which produces pre-pulses in the temporal domain, as seen in Fig. 5(d). This residual phase is quite common in prism compressors (which are known to overcompensate the TOD of materials) and in principle can be minimized, for a given compressor position, by adjusting the TOD-compensating mirrors within the laser system[16]. The optimum pulse duration is 27.5±0.2 fs (FWHM).

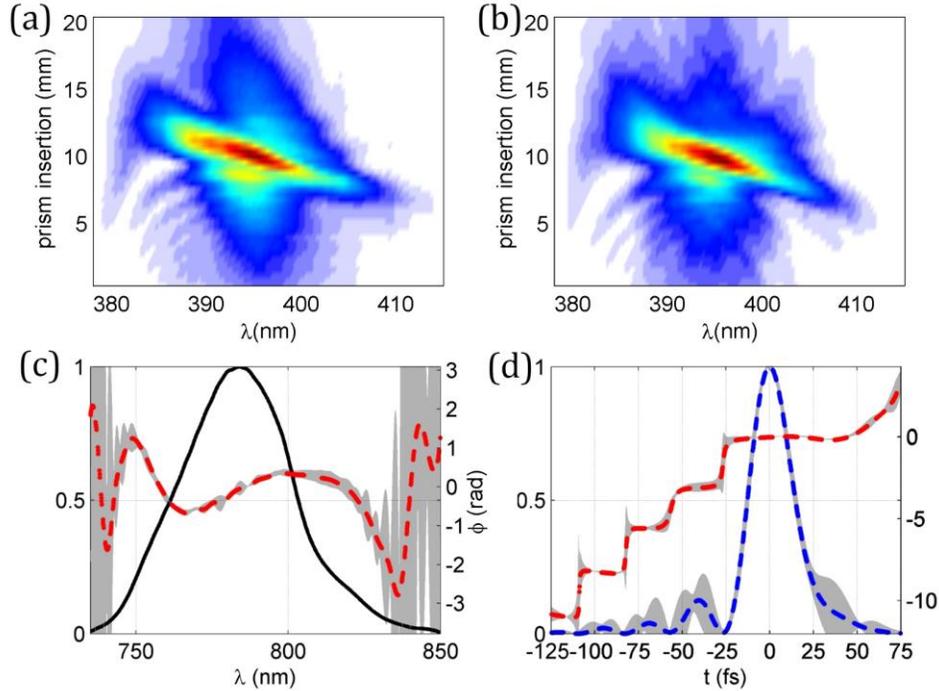

**Figure 5**. **Results of the CPA laser pulses measured with self-calibrated d-scan**. (**a**) Experimental and (**b**) retrieved self-calibrated d-scan traces. (**c**) Spectrum of the pulse (solid black curve) and retrieved phase (dashed red curve), (**d**) temporal intensity (dashed blue curve) and retrieved phase (dashed red curve). Grey shaded areas denote the error (magnified 10 times) in the retrievals, calculated as the standard deviation over 5 independent measurements.

To conclude, we presented and demonstrated a self-calibrating d-scan technique, which allows measuring ultrashort pulses without prior knowledge of the dispersion introduced in the scan. Numerical simulations and experimental measurements of a few-cycle laser oscillator using a chirped mirror and glass wedge compressor showed that the retrieval algorithm calculates both the spectral phase of the pulse and the differential dispersion of the compressor. The technique was then applied to pulses produced by a CPA laser, where the dispersion scan was introducing by varying the internal pulse compressor of the system. Together with the pulse characterization, the self-calibrating d-scan method simultaneously calibrated the amount of dispersion introduced per step of the compressor and optimized the compression to find the shortest pulse duration generated from the system, in this case 27.5 fs. The present work demonstrates that *the d-scan technique can be extended to any ultrashort pulse duration range*, and applied to any CPA or OPCPA system, as well as to any laser system coupled to an arbitrary compressor. Such systems can therefore be self-diagnosed with the help of only a nonlinear medium and a spectrometer. The self-calibrating d-scan also allows performing on-target diagnostics by placing the nonlinear medium at the target location and measuring the d-scan trace at that point, without the need of additional calibrations or diagnostics, hence providing complete temporal information about the pulse effectively delivered by the laser system to the experimental target.

# Methods

The essential information required to understand this work has already been presented along the manuscript, in particular regarding the experimental implementation of the d-scan technique and the corresponding retrieval algorithm used to analyze the data, since it constitutes the core of the work. Here, we provide further details on the experiments and the algorithm employed for the self-calibrating d-scan retrieval.

**Experimental details**. All the experiments were performed using different outputs of the same Ti:Sapphire laser system from Femtolasers Produktions GmbH. The ultrafast oscillator (Femtosource Rainbow CEP) output is carrier-envelope phase (CEP) stabilized and delivers pulses with a Fourier-transform limit of around 7 fs at a repetition rate of 80 MHz, with a central wavelength of around 800 nm and an energy per pulse of up to 2.5 nJ. The CPA amplified output (FemtoPower Compact PRO CEP) provides 1 kHz, ~1 mJ pulses with a Fourier-transform limit of ~22 fs.

In the experiments, the SHG signal was optimized using the criterion of maximum signal and we verified that it was spatially homogeneous by laterally translating the collection fibre of the spectrometer.

In the particular case of a prism compressor, it is important to avoid nonlinear effects due to propagation, e.g. in the bulk of the prisms, which may vary along the d-scan measurement, since the pulse intensity and path length within the bulk depend on the prism insertion. This is a well-known issue with prism compressors, where maximum compression can occur too close to, or even inside, the last prism thus giving rise to unwanted self-phase modulation effects[17]. Normally, this will not happen when using grating compressors, where the corresponding B-integral is much lower. To avoid this problem, the output energy of the amplifier was reduced by decreasing the energy of the CPA pump laser together with further attenuation using beamsplitters placed after the amplification and before the prism compressor. This way we obtained a pulse energy of 20 μJ at the output. We checked that the d-scan trace did not vary when reducing the energy by a factor of two (i.e. 10 μJ) and that the fundamental spectrum was also unmodified during the scan, which is indicative of the absence of nonlinear effects. Note that this energy reduction is not necessary when using lower energy laser systems or those equipped with a diffraction grating compressor, where the nonlinear phase accumulated during propagation in the compressor is in principle very low.

**Self-calibrating d-scan and retrieval algorithm**. The self-calibrating d-scan is used to measure the temporal (and spectral) amplitude and phase of the laser pulses. The electric field of the pulse is defined in the spectral domain as $E(\omega) = A(\omega)\exp\{i\varphi(\omega)\}$, where $\omega$ is the frequency, and $A(\omega)$, $\varphi(\omega)$ are the spectral amplitude and phase, respectively. The amplitude is directly obtained from the power spectral density defined as $S(\omega) = |A(\omega)|^2$, which is calculated from the signal measured with the spectrometer $S(\lambda)$, including the correction scaling factor of $\lambda^2$ that ensures the conservation of the integral of the spectral distribution. During the dispersion scan, the compressor adds dispersion to the pulse, while the SHG signal is measured. The uncalibrated compressor (e.g. based on prisms or gratings), or the bulk material (e.g. wedges), introduces a phase $\phi(\omega)$ that depends on the compressor position (e.g. the prism or grating translation), or the inserted thickness of material (e.g. the wedge insertion), represented by $z$. This phase is modeled by a truncated Taylor series as given by

$$\frac{\phi(\omega)}{z} = \phi_0 + \phi_1 \cdot (\omega - \omega_0) + \frac{1}{2}GDD \cdot (\omega - \omega_0)^2 + \frac{1}{6}TOD \cdot (\omega - \omega_0)^3 + \cdots \qquad (1)$$

where $\omega_0$ is the central frequency. As mentioned previously, the first two terms of the series do not affect the d-scan trace and are disregarded, while the coefficients GDD, TOD, etc., are calibrated from the algorithm.

The measured SHG d-scan trace as a function of the frequency and the compressor $z$ position can be expressed as

$$S_{meas}(\omega, z) = |\mathcal{F}[(\mathcal{F}^{-1}\{A(\omega)\exp[\varphi(\omega)]\exp[i\phi(\omega) \cdot z]\})^2]|^2 \qquad (2)$$

where $\mathcal{F}$ denotes the Fourier transform. The phase of the pulse, $\varphi(\omega)$, is sampled over a number of points (typically 32-48 points) that are evaluated as unknown variables. Then, the phase is interpolated to the frequency axis so that it is used within the algorithm to retrieve the pulse in the temporal domain. The unknown parameters of the compressor (GDD, TOD, etc.) and the sampled phase of the pulse therefore form a set of multiple unknown variables.

The calculation of those variables is done with a multi-variable optimization algorithm (e.g., Nelder-Mead Simplex[2] or, in this case, the Levenberg-Marquardt algorithm), in which the simulated d-scan trace, $S_{simul}(\omega, z)$, is computed and compared to the measured trace, $S_{meas}(\omega, z)$. In this iterative algorithm, the difference between the traces is minimized (using a merit function[2]) considering all the unknowns as a whole (sampled spectral phase and the dispersion of the compressor), therefore retrieving the pulse and calibrating the compressor at the same time. The pulse is calculated for every position of the compressor, $z$, so we can obtain the pulse measurement for any particular position, including for optimum compression (position that gives the shortest pulse duration and highest peak power).

We have tested our algorithm with experimental measurements, and we have systematically found the same solution, within the tolerance error, even when using initial conditions (guesses) for the spectral phase of the pulse and the compressor dispersion that deviate significantly from the solution. Also, there is no need to calibrate the amplitude of the measured SHG signal in the d-scan trace, since this is done by the algorithm using a frequency-dependent error function, defined as[2]

$$\mu_i = \frac{\sum_j S_{meas}(\omega_i, z_j) S_{simul}(\omega_i, z_j)}{\sum_j S_{simul}(\omega_i, z_j)}, \quad (3)$$

which corresponds to the merit function (overall error) given by

$$G = \sqrt{\frac{1}{N_i N_j} \sum_{i,j} \left(S_{meas}(\omega_i, z_j) - \mu_i S_{simul}(\omega_i, z_j)\right)^2}. \quad (4)$$

The iteration is stopped when the convergence between measured and retrieved traces is reached, which is defined as the merit function not improving for a number of consecutive iterations (for example, when the relative variation of $G$ is $< 10^{-5}$ during 10 consecutive iterations).

**Data availability**. The datasets generated and/or analyzed during the current study are available from the corresponding author on reasonable request.

## Acknowledgements

We thank Francisco Silva for very fruitful suggestions and discussions about the method. We also acknowledge funding from Junta de Castilla y León (SA116U13, SA046U16) and Spanish MINECO (FIS2013-44174-P, FIS2015-71933-REDT, FIS2017-87970-R), and Fundação para a Ciência e Tecnologia (FCT), Portugal, via grants SFRH/BPD/88424/2012, SFRH/BSAB/105974/2015, M-ERA-NET2/0002/2016, UID/NAN/50024/2013, NORTE-07-0124-FEDER-000070, NORTE-01-0145-FEDER-022096, and the Network of Extreme Conditions Laboratories (NECL), co-funded by COMPETE and FEDER.

## Author contributions statement

B.A. proposed the self-calibrating algorithm. B.A. and H.C. conducted the experiments. I.J.S and H.C. supervised the project. B.A. performed the simulations and analyzed the experimental results. All the authors were involved in the interpretation of the results and contributed to the manuscript preparation.